# MR Optimized Reconstruction of Simultaneous Multi-Slice Imaging Using Diffusion Model

Ting Zhao[1,2], Zhuoxu Cui[1], Sen Jia[1], Qingyong Zhu[1], Congcong Liu[1], Yihang Zhou[1], Yanjie Zhu[1], Dong Liang[1], and Haifeng Wang[1]

[1]Shenzhen Institutes of Advanced Technology, Chinese Academy of Sciences, Shenzhen, China, [2]University of Chinese Academy of Sciences, Beijing, China

## Synopsis

**Motivation:** Diffusion model has been applied to MRI reconstruction, including single and multi-coil acquisition of MRI data. Simultaneous multi-slice imaging (SMS), as a method for accelerating MR acquisition, significantly reduces scanning time, but further optimization of reconstruction results is still possible.

**Goal(s):** In order to optimize the reconstruction of SMS, we proposed a method to use diffusion model based on slice-GRAPPA and SPIRiT method.

**Approach:** Specifically, our method characterizes the prior distribution of SMS data by score matching and characterizes the k-space redundant prior between coils and slices based on self-consistency.

**Results:** With the utilization of diffusion model, we achieved better reconstruction results.

**Impact:** The application of diffusion model can further reduce the scanning time of MRI without compromising image quality, making it more advantageous for clinical application.

## Introduction

There are many methods for SMS, including SENSE-based SMS[1], 2D CAIPIRINHA[2], and SENSE-GRAPPA and Slice-GRAPPA based on GRAPPA[3]. These methods are typically used to deal with inter-slice aliasing, and other parallel imaging methods are needed to deal with intra-slice undersampling-induced aliasing. In addition, research has also been conducted to add sparse constraint optimization to SMS reconstruction based on the above methods[4]. However, the current methods still cannot achieve satisfactory reconstruction results at high acceleration factors. With the proposal of score based generative models Song et al.[5], accurate estimation of the image prior distribution $p(x)$ becomes possible. Currently, diffusion models have achieved satisfactory results in multi-coil MR reconstruction, making it possible to further apply diffusion models to SMS for higher acceleration factors and better reconstruction results. Inspired by SPIRiT-diffusion[6], our proposed method incorporates Slice-GRAPPA to achieve diffusion of SMS data, optimizing the reconstruction results.

## Method

The reconstruction of SMS using Slice-GRAPPA and SPIRiT can be represented using the following objective function:

$$\arg\min_x ||(H-I)\tilde{x}||^2 + \lambda ||Dx - y||^2$$

Where $x = \begin{bmatrix} x_1 \\ x_2 \\ x_3 \end{bmatrix}$, meaning the multi-slice image, $\tilde{x}$ is the corresponding k-space data,

$$H = \begin{bmatrix} K_1 \\ K_2 \\ K_3 \end{bmatrix} \begin{bmatrix} I & I & I \end{bmatrix} \begin{bmatrix} G_1 & & \\ & G_2 & \\ & & G_3 \end{bmatrix}$$

meaning using SPIRiT operators $G_i$ that convolve the entire undersampling k-space to interpolate in the missing k-space data, and using slice-GRAPPA operators $K_i$ to separate different slices. $D$ is the sampling matrix and $y$ is the SMS data. Eq. 1 can be solved using the following iterative solution algorithm:

$$x_i = x_{i+1} + 2\eta_{i+1}\mathcal{F}^{-1}((H-I)^*(H-I)\mathcal{F}(x_{i+1})) + \lambda_{i+1}(Dx - y)$$

Taking Eq. 2 as the iteration of the reverse diffusion process, the corresponding forward diffusion process can be defined. To solve the calculation of covariance of the perturbation kernel in the diffusion process, we add a operator $\mathcal{T}$ to the standard Wiener process to enforce the noises in diffusion process satisfy the self-consist operator $H$, and it can be presented as follows:

$$\mathcal{T}(z) = \arg\min_z ||(H-I)\tilde{z}||^2$$

Here, z is the Gaussian noise added in the diffusion process, $\tilde{z}$ is the corresponding k-space. The reverse diffusion process is defined as:

$$dx = (\frac{\eta(t)}{2}\Psi(x) - \beta(t)\mathcal{T}(\nabla_x log p_x(x|y)))dt + \sqrt{\beta(t)}\mathcal{T}dw$$

where $\Psi(x) = \mathcal{F}^{-1}((H-I)^*(H-I)\mathcal{F}(x))$, $\frac{\eta(t)}{2}\Psi(x)$ is the drift coefficient of $x(t)$ and $\sqrt{\beta(t)}\mathcal{T}dw$ is the diffusion coefficient of $x(t)$. The corresponding forward diffusion process becomes:

$$dx = \frac{\eta(t)}{2}\Psi(x)dt + \sqrt{\beta(t)}\mathcal{T}dw$$

The perturbation kernel of Slice-Diffusion can be derived as:

$$p_{0t}(x(t)|x(0)) = \mathcal{N}(x(t); x(0), \sigma^2 \mathcal{T})$$

then, the score model can be trained using U-Net via:

$$\theta^* = \arg\min_{\theta} \mathbb{E}_t\{\lambda(t)\mathbb{E}_{x(0)}\mathbb{E}_{x(t)|x(0)}\left[||\theta\mathcal{T}(s_\theta(x(t),t)) + z||^2\right]\}$$

## Results

We conducted a retrospective experiment on the fastMRI dataset, simulating SMS data with a multi-slice factor of 3. To reduce the noise amplification, CAIPIRINHA was utilized with a phase increment of 2π/3 that provides 1/3 FOV shifts in between the adjacent slices. Fig 2 shows results from retrospectively 3-fold SMS and 3-fold in-plane accelerated images. Our proposed method further optimized the reconstruction results of SMS. Fig 3 shows results from 3-fold SMS and 10-fold in-plane accelerated images. Under such extreme undersampling conditions, Our proposed method achieved fairly good reconstruction results, although there were some detail errors due to highly undersampling.

## Acknowledgements

Ting Zhao and Zhuoxu Cui contributed equally to this work. This work was partially supported by the National Natural Science Foundation of China (62271474), the National Key R&D Program of China (2023YFB3811400), the High-level Talent Program in Pearl River Talent Plan of Guangdong Province (2019QN01Y986) and the Shenzhen Science and Technology Program (KQTD20180413181834876 and JCYJ20210324115810030).

## References

1. Hamilton, Jesse, Dominique Franson, and Nicole Seiberlich. "Recent advances in parallel imaging for MRI." Progress in nuclear magnetic resonance spectroscopy 101 (2017): 71-95.

2. Breuer, Felix A., et al. "Controlled aliasing in volumetric parallel imaging (2D CAIPIRINHA)." Magnetic Resonance in Medicine: An Official Journal of the International Society for Magnetic Resonance in Medicine 55.3 (2006): 549-556.

3. Setsompop, Kawin, et al. "Blipped-controlled aliasing in parallel imaging for simultaneous multislice echo planar imaging with reduced g-factor penalty." Magnetic resonance in medicine 67.5 (2012): 1210-1224.

4. Demirel, Ömer Burak, et al. "Improved Simultaneous Multi-slice imaging with Composition of k-space Interpolations (SMS-COOKIE) for myocardial T1 mapping." PloS one 18.7 (2023): e0283972.

5. Song, Yang, et al. "Score-based generative modeling through stochastic differential equations." arXiv preprint arXiv:2011.13456 (2020).

6. Cao, Chentao, et al. "SPIRiT-Diffusion: SPIRiT-driven Score-Based Generative Modeling for Vessel Wall imaging." arXiv preprint arXiv:2212.11274 (2022).

## Figures

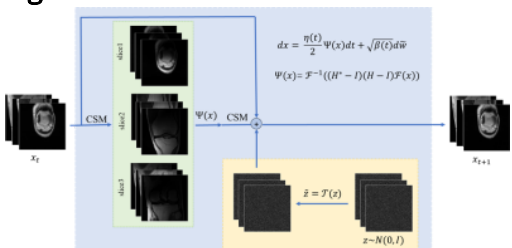

Fig.1 Diffusion process of proposed method. H is the consistency operator.

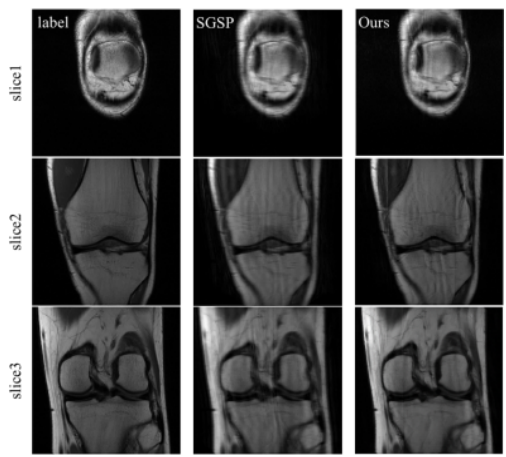

Fig.2 Retrospectively 3-fold SMS and 3-fold in-plane accelerated results. 32 ACS lines were used. SGSP means the slice-GRAPPA and SPIRiT results.

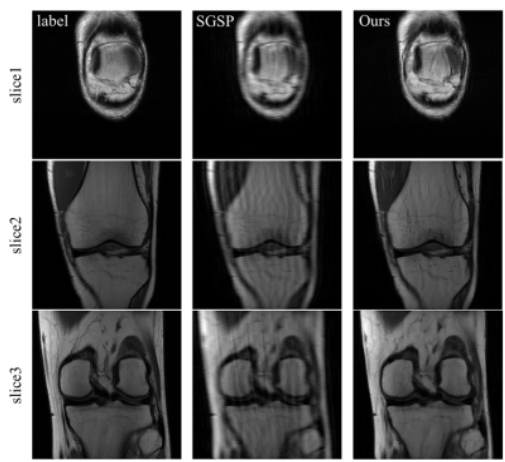

Fig.3 Retrospectively 3-fold SMS and 10-fold in-plane accelerated results. 32 ACS lines were used. SGSP means the slice-GRAPPA and SPIRiT results.

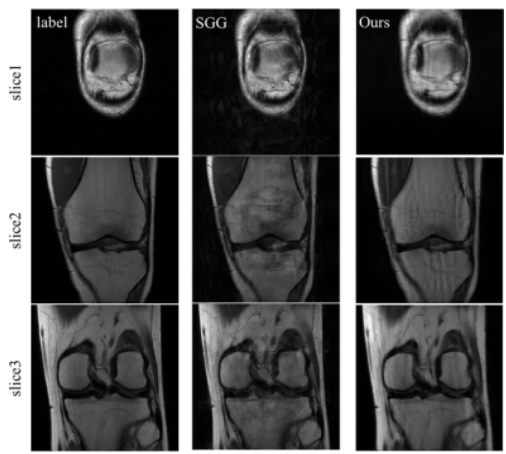

Fig.4 Retrospectively 3-fold SMS and 3-fold in-plane accelerated results. 32 ACS lines were used. SGG means the slice-GRAPPA and GRAPPA results.